\documentclass[12pt,twoside]{article}\usepackage[super,sort&compress,comma]{natbib}
\usepackage{fancyhdr}

\usepackage[section]{placeins}   %
\usepackage{graphicx}
\usepackage{subfigure}

\makeatletter \renewcommand\@biblabel[1]{$^{#1}$} \makeatother
\newcommand{\cen}[1]{\begin{center} #1 \end{center}}

\usepackage[mathlines]{lineno}

\usepackage{hyperref}
\hypersetup{ colorlinks,
    citecolor=blue,
    filecolor=blue,
    linkcolor=blue,
    urlcolor=blue
}

\usepackage{xcolor}
\definecolor{gray}{rgb}{0.6,0.6,0.6}
\definecolor{red}{rgb}{0.85,0,0}
\definecolor{green}{rgb}{0,0.85,0}
\definecolor{blue}{rgb}{0,0,0.85}
\definecolor{beige}{rgb}{0.92,0.87,0.78}
\usepackage[all]{hypcap}    
\begin{document}

\cen{\sf {\Large {\bfseries Detectability assessment of an x-ray imaging system using the nodes in a wavelet packet decomposition of a star-bar object} \\  
\vspace*{10mm}
Antonio Gonz\'alez-L\'opez}\\
Hospital Universitario Virgen de la Arrixaca, ctra. Madrid-Cartagena, 30120 El Palmar (Murcia), Spain}
·
\vspace{5mm}\\
Version typeset \today\\
%
\pagenumbering{roman}
\setcounter{page}{1}
\pagestyle{plain}
Author to whom correspondence should be addressed. email: antonio.gonzalez7@carm.es\\
\begin{abstract}
\noindent {\bf Purpose: Using linear transformation of the data allows studying detectability of an imaging system on a large number of signals. An appropriate transformation will produce a set of signals with different contrast and different frequency contents. In this work both strategies are explored to present a task-based test for the detectability of an x-ray imaging system.} \\
{\bf Methods: Images of a new star-bar phantom are acquired with different entrance air KERMA and with different beam qualities. Then, after a wavelet packet is applied to both input and output of the system, conventional statistical decision theory is applied to determine detectability on the different images or nodes resulting from the transformation. A non-prewhitening matching filter is applied to the data in the spatial domain, and ROC analysis is carried out in each of the nodes.} \\
{\bf Results: AUC maps resulting from the analysis present the area under the ROC curve over the whole 2D frequency space for the different doses and beam qualities. Also, AUC curves, obtained by radially averaging the AUC maps allows comparing detectability of the different techniques as a function of the frequency in one only figure. The results obtained show differences between images acquired with different doses for each of the beam qualities analyzed.} \\
{\bf Conclusions: Combining a star-bar as test object, a wavelet packet as linear transformation, and ROC analysis results in an appropriate task-based test for detectability performance of an imaging system. The test presented in this work allows quantification of system detectability as a function of the 2D frequency interval of the signal to detect. It also allows calculation of detectability differences between different acquisition techniques and beam qualities.} 
\end{abstract}
\noindent{\it Keywords}: Detectability, wavelet packet, NPW observer, star-bar phantom

\newpage     

\newpage

\setlength{\baselineskip}{0.7cm}      

\pagenumbering{arabic}
\setcounter{page}{1}
\pagestyle{fancy}
\section{Introduction}
Basic performance assessment of imaging detectors consists in measuring physical parameters such as MTF, NPS and DQE\cite{Oliveira2019,Despres2018,Strudley2015,Vassileva2010,Samei2011}. Then, these results are used to verify technical compliance with system specifications in terms of physical parameters of the imaging sensor such as spatial resolution and noise. However, these parameters do not provide any definitive way of rating the image quality of a system, since they are largely task independent and any general definition of image quality must be based in the effectiveness with which the image can be used for its intended task\cite{ICRU1995}.

Nowadays, image quality assessment in terms of task-based tests is being developed\cite{Samei2019,Hernandez-Giron2015,VanPeteghem2016,Han2018,Russ2016,Solomon2015,MakiBloomquist2014}. A task-based assessment aims to quantify the ability of a human observer to perform a given task. The purpose of these tests is to determine the clinical performance of the imaging system, in such a way that the evaluation results are more consistent with its diagnostic capabilities.
 
The main goal of quality assurance is to ensure the accuracy of the diagnosis while minimizing the radiation dose\cite{Uneri2018,DelasHerasGala2018,Jang2018,Maldera2017,Dance2016,Eck2015,Vano2013,Luck2013,Yan2012}. From a diagnostic perspective, one of the most important tasks determining image quality is detectability. Detectability of an object in an image depends on the radiation beam used for image acquisition, some physical properties of the imaging system such as noise and spatial resolution, some characteristics of the object being imaged like its size and contrast and on the task function and the observer model used\cite{Gang2014,ICRU1995}. Because of the impact of detectability and dose on the quality of diagnostic procedures, determining metrics that describe the relationship between them must be a fundamental objective of quality assurance.

The image $g_s$ of an object $f_s$ acquired with an imaging system represents the output of the system to an input. In the case of a linear system, the output $g_s$ can be calculated from the input $f_s$, a linear operator $H$ and noise $n$\cite{ICRU1995} as $g_s=Hf_s+n$. If no signal is present in the image, the output will be $g_{as}=Hf_{as}+n=n$, where $f_{as}\equiv 0$ stands for the absence of signal. If no previous information on the presence of signal is available, output image $g$ can be expressed as a function of noise and the unknown input,
\begin{equation}\label{eq:transfer}
g=Hf+n.
\end{equation}

Statistical decision theory is applied to study the detectability performance of an imaging system by analyzing images containing a known signal and images containing only background. This methodology allows comparison of the quality of images acquired with different techniques and different doses.

A fundamental part of detectability analysis is the signal used. In particular, the study will be highly improved is a set of signals of different degree of detectability is available. A method for obtaining a set of input signals is decomposing the object in a linear combination of signals. Then, if the system is linear, the output image will be the linear combination of the outputs to each of these components. In this way, by decomposing both input and output, detectability can be studied on each of the input components that will constitute the set of test input signals.

A further refinement is using a linear transformation of the data. If a linear transformation $W$ is applied to both input $f$ and output $g$, in a linear system detectability of the transformed object (new input) $Wf$ can be studied by analyzing the transformed image (new output) $Wg$. In this case, equation \ref{eq:transfer} transforms into
\begin{equation}\label{eq:wtransfer}
Wg=WHf+Wn.
\end{equation} 
This is the idea followed in this work, and the linear transformation used is a wavelet packet. A wavelet packet decomposes an image in smaller images. The interest of these new images is that their frequency contents are restricted to a small area of the frequency space. In addition, the contrast of the new images is smaller, making harder their detection.

According to ICRU\cite{ICRU1995}, for an ideal Bayesian observer, decision between two hypothesis, signal present ($H_1$) and signal absent ($H_2$) should be based in the likelihood ratio $L$,
\begin{equation}\label{eq:likeratio1}
L=\frac{p(Wg|H_2)}{p(Wg|H_1)}
\end{equation}

If distribution for noise $n$ is Gaussian then $Wn$ is also Gaussian and $L$ can be expressed as\cite{Fukunaga1990}
\begin{equation}\label{eq:likeratio2}
L=(W(f_{as}-Hf_s))^t C_{Wn}^{-1}Wg,
\end{equation}
where $f_{as}-Hf_s=-Hf_s$ is the difference between the input signals under the two hypothesis\cite{ICRU1995}, $C_{Wn}$ is the covariance matrix for transformed noise and $t$ indicates the transpose. 

If no prewhitening is carried out, $C_{Wn}^{-1}$ is removed from the precedent equation and the decision variable transforms into the non-prewhitening NPW matching filter index,
\begin{equation}\label{eq:lr_npw}
L_{NPW}=(W(f_{as}-Hf_s))^t Wg.
\end{equation}
\section{Material and methods}
\begin{figure}
\centering
{\includegraphics[width=0.7\textwidth]{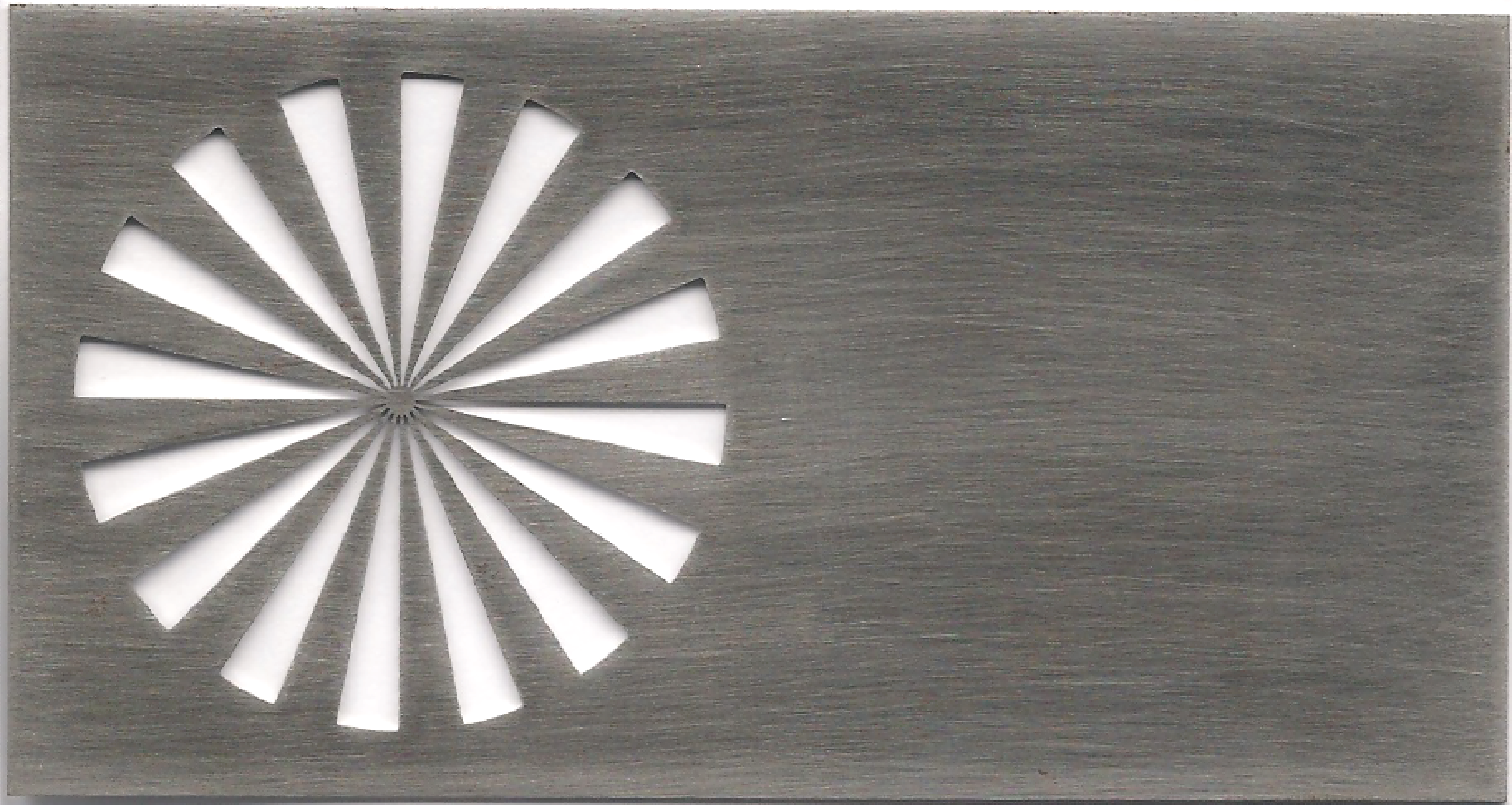}}
\caption{The star-bar phantom used in this work is a plate of $115 mm\times 60mm$ with $1\:mm$ thickness and manufactured in stainless steel.}
\label{fig:maniq}
\end{figure}

The x-ray beam used for image acquisition was generated in an Ysios system from Siemens. Two beam qualities were used RQA 3 and RQA 5. Also, for each beam quality two entrance air KERMA were utilized, 1.79 and 17.9 $\mu Gy$ for RQA 3 and 1.76 and 7.99 $\mu Gy$ for RQA 5. For each combination of beam quality and air KERMA, 25 images were acquired. The imaging detector was a PIXIUM 3543 pR from Trixell (CsI coupled to TFT matrix of $a$Si) with a pixel spacing of $0.144\, mm \times 0.144\, mm$ and the star-bar phantom was manufactured in stainless steel (figure \ref{fig:maniq}) and consisted of a $115 mm\times 60mm$ plate of $1\:mm$ thickness with triangular holes. These holes produce sixteen bar pairs with a variable period that ranges between $9.8\:mm$ and $0.4\:mm$. 

\begin{figure}
\centering
\subfigure[$g_s$]
{\includegraphics[width=0.325\textwidth]{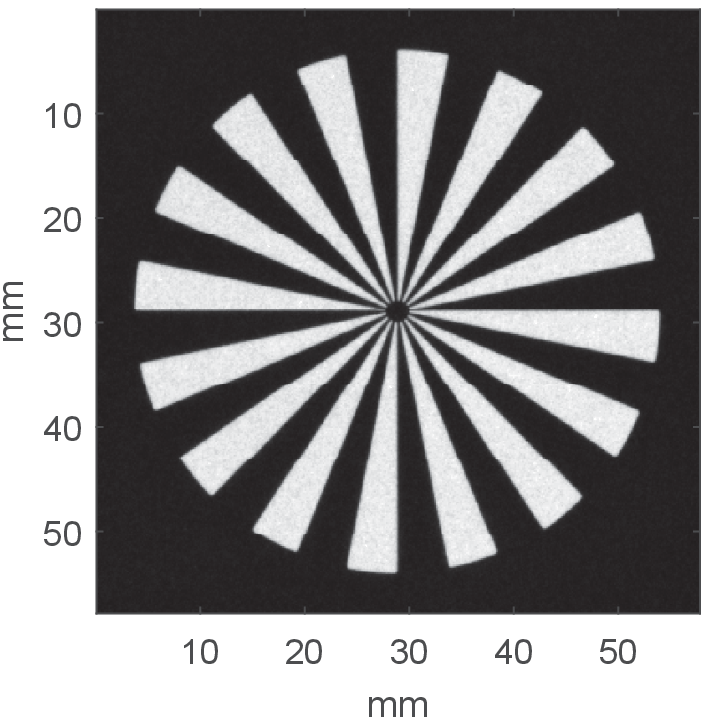}\label{fig:g_s}}
\subfigure[$Hf_s$]
{\includegraphics[width=0.325\textwidth]{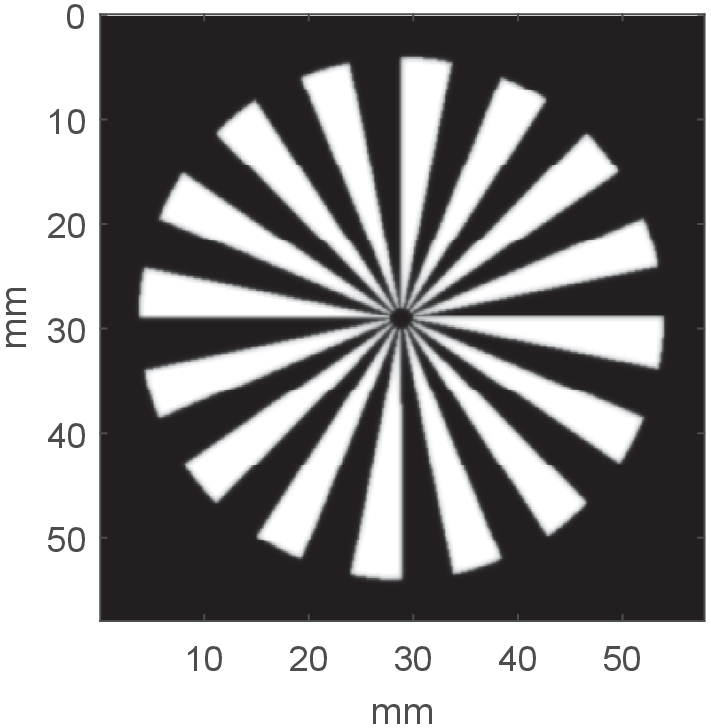}\label{fig:Hf_s}}
\subfigure[$g_{as}$]
{\includegraphics[width=0.325\textwidth]{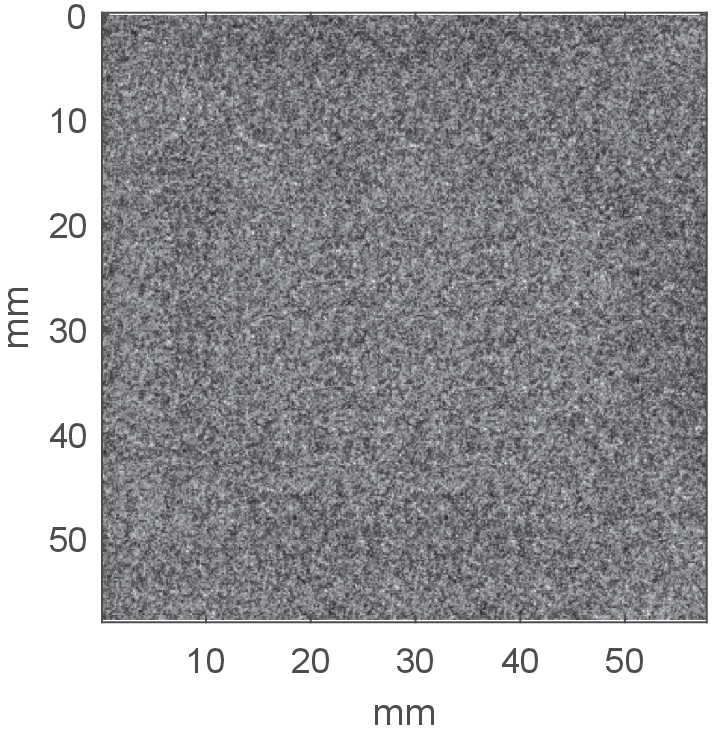}\label{fig:g_as}}
\caption{A sample of the images used in this study. The image of the star-bar represents the system output $g_s$ to input signal $f_s$, the template $Hf_s$ is obtained as the convolution of the system PSF with a synthetic image of the star-bar phantom and noise in a uniform area of the image is used as the absence of signal $g_{as}$.}
\label{fig:sig_orig}
\end{figure}

The method for studying detectability presented in this work follows two main steps. 
\begin{enumerate}
\item First, each of three large images is decomposed into a number of small images. The first of these images contains the signal and is the star-bar image acquired with the imaging system (figure \ref{fig:g_s}); the second image is a template obtained as the convolution of an ideal star-bar image $f_s$ and the point spread function of the system $H$ (figure \ref{fig:Hf_s}); the third image is a noise image with no signal (figure \ref{fig:g_as}). 
\item Second, statistical decision theory using a computer observer model is applied to each of the small images in which $f_s$ is decomposed to study the system capability to detect them.
\end{enumerate}

For the decomposition of the signals a wavelet packet\cite{STEPHANE2009377} or subband tree is used. For detectability a non-prewhitening observer and ROC analysis are used.

\subsection{Wavelet decomposition}
\begin{figure}
\centering
{\includegraphics[width=0.8\textwidth]{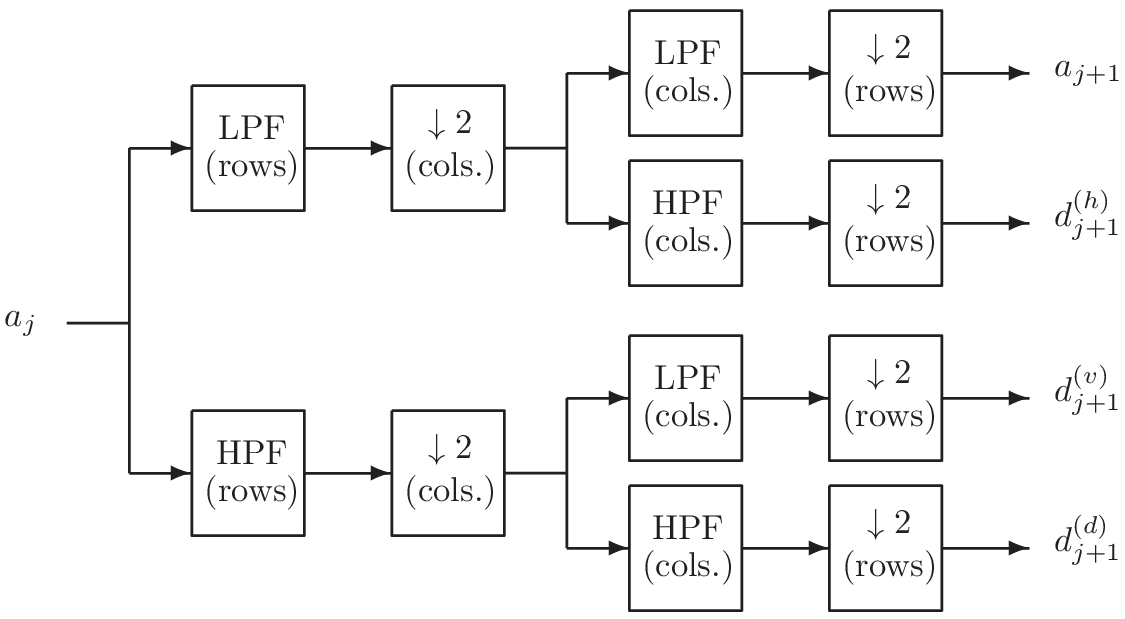}}
\caption{Transformations applied to one of the images at level $j$ to obtain 4 images at level $j+1$ following a wavelet packet decomposition.}
\label{fig:wltx2d}
\end{figure}

A wavelet packet is a wavelet transform organized in a number $L$ of levels. Level 0 correspond to the original image to decompose and, at each of the successive levels $j$, the images in the preceding level $j-1$ are filtered by applying a combination of low-pass filters (LPF) and high-pass filters (HPF) to its rows and columns as shown in figure \ref{fig:wltx2d} followed by a factor 2 downsampling in both rows and columns. In this way, for each of the images at level $j-1$ 4 images or nodes are produced at level $j$. The final number of images is $4^L$. However, due to downsampling, each of the images produced in level $j$ has $1/4$ the size of the image at level $j-1$.

\begin{figure}
\centering
{\includegraphics[width=0.85\textwidth]{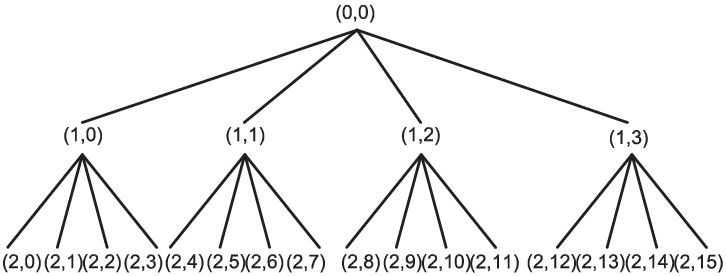}}
\caption{Two-levels wavelet packet. The original image is represented by node $(0,0)$ and each image or node $(j,k),\: j=1,2;\: k=0,1,2,3$ is obtained from one of the nodes in the precedent level $j-1$ following the scheme shown in figure \ref{fig:wltx2d}.}
\label{fig:packet_treet}
\end{figure}

Figure \ref{fig:packet_treet} shows a schematic of a 2-levels wavelet packet ($L=2$). Each image in a wavelet packet decomposition is called a node, node $(0,0)$ being the original image. Nodes at level $j+1$ $(j+1,0)$, $(j+1,1)$, $(j+1,2)$ and $(j+1,3)$ in figure \ref{fig:packet_treet} correspond to images $a_{j+1}$, $d_{j+1}^{(h)}$, $d_{j+1}^{(v)}$, $d_{j+1}^{(d)}$ in figure \ref{fig:wltx2d} respectively.

\begin{figure}
\centering
\subfigure[Two levels]
{\includegraphics[width=0.46\textwidth]{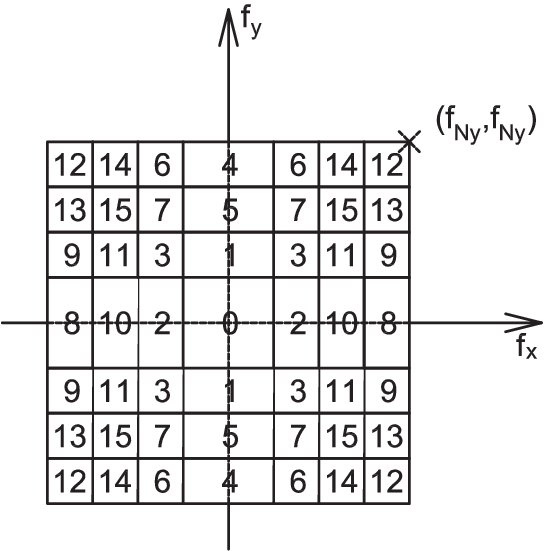}\label{fig:freccont2lev}}
\subfigure[Seven levels]
{\includegraphics[width=0.5\textwidth]{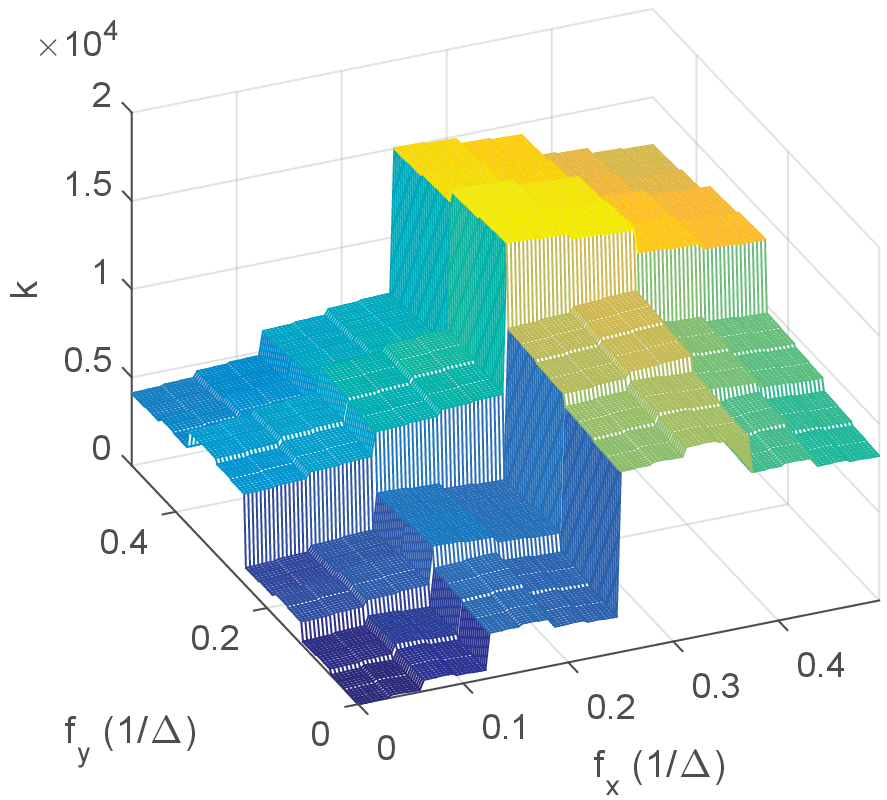}\label{fig:freccont7lev}}
\caption{(a) Frequency contents of images or nodes $(2,k),\:k=0,1,..,15$ resulting from the 2-levels wavelet packet in figure \ref{fig:packet_treet}. (b) Frequency contents of nodes $(7,k)$ for the 7-levels wavelet packet used in this work.}
\label{fig:freccont}
\end{figure}

A wavelet packet applies a series of band pass filters in such a way that each of the images or nodes at a given level encapsulates the frequency content of the original image in a rectangular area of the frequency space. Figure \ref{fig:freccont2lev} shows the frequency contents for each of the nodes resulting from decomposition in figure \ref{fig:packet_treet}. For instance, node $(2,13)$ will have its energy concentrated in the frequency interval $(3/4f_{Ny},f_{Ny})\times(1/2f_{Ny},3/4f_{Ny})$. Figure \ref{fig:freccont7lev} describes the frequency contents for each of the nodes in the 7-levels wavelet packets used in this work.

\begin{figure}
\centering
\subfigure[Haar]
{\includegraphics[width=0.49\textwidth]{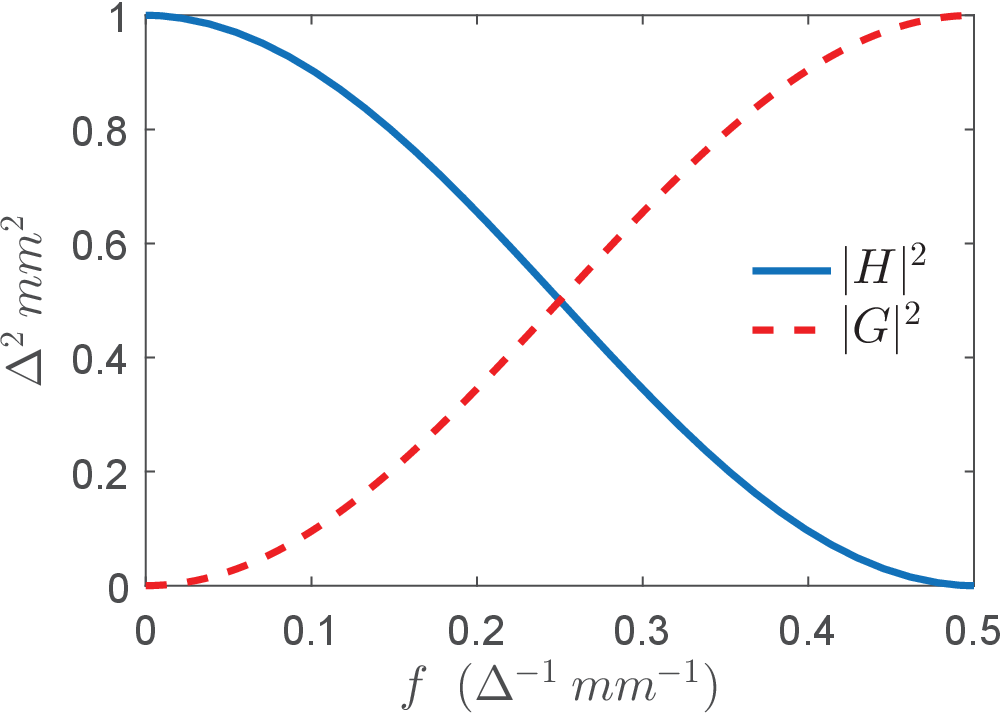}\label{fig:fils_haar}}
\subfigure[Sym16]
{\includegraphics[width=0.49\textwidth]{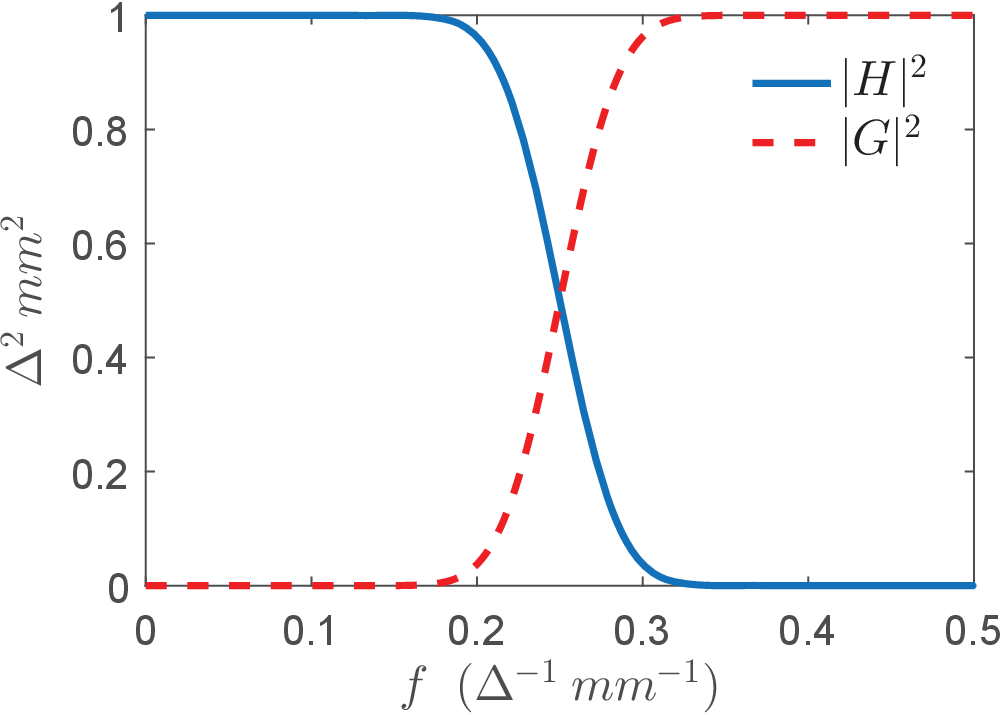}\label{fig:fils_sym16}}
\caption{Frequency response of low-pass filter $H$ and high-pass filter $G$ used in figure \ref{fig:wltx2d} for two different wavelets. Responses of sym16 wavelets filters are sharper than those of Haar filters making them closer to ideal low-pass and high-pass filters.}
\label{fig:fils}
\end{figure}

Figure \ref{fig:fils} shows two different pairs of wavelet filters. It can be seen how high and low-pass filters for wavelet sym16 have a sharper response than those of the Haar wavelet. This means that the frequency division carried out by the filter banks of figure \ref{fig:wltx2d} is more efficient. For this reason, sym16 has been the wavelet used in this work. It should be noted that increasing the complexity of these wavelets improves filters performance but increases calculation times.

\subsection{Detectability}
If node $(L,k)$ of the wavelet packet of signal $s$ is written as $Ws(L,k)$, for the linear transform carried out by the wavelet packet, detectability indexes for test signal $Wf_s(L,k)$ can be calculated using equation \ref{eq:lr_npw} as the non-prewhitening matching filter\cite{ICRU1995} for the signal-present image $Wg_s$,
\begin{equation}\label{eq:npw1}
d_s(L,k)=\sum_{l=0}^{N-1}(WHf_s(L,k)-Wf_{as}(L,k))Wg_s
\end{equation}
and for the signal-absent image $Wg_{as}$,
\begin{equation}\label{eq:npw2}
d_{as}(L,k)=\sum_{l=0}^{N-1}(WHf_s(L,k)-Wf_{as}(L,k))Wg_{as},
\end{equation}
where $H$ is the system PSF and $f_{as}\equiv 0$. With these indexes, ROC curves can be calculated by comparing each $d$ value with a threshold to determine if the signal is present or not. Using indexes obtained from one node of the signal and one node of noise will give us a point of the ROC curve. Then, by varying the value of the threshold the entire ROC curve for that node will be obtained.

It should be noted that matching filtering is carried out in the spatial domain. For this reason, before applying equation \ref{eq:npw1} an accurate registration of $g_s$ and $Hf_s$ must be carried out. 

\begin{figure}
\centering
{\includegraphics[width=1\textwidth]{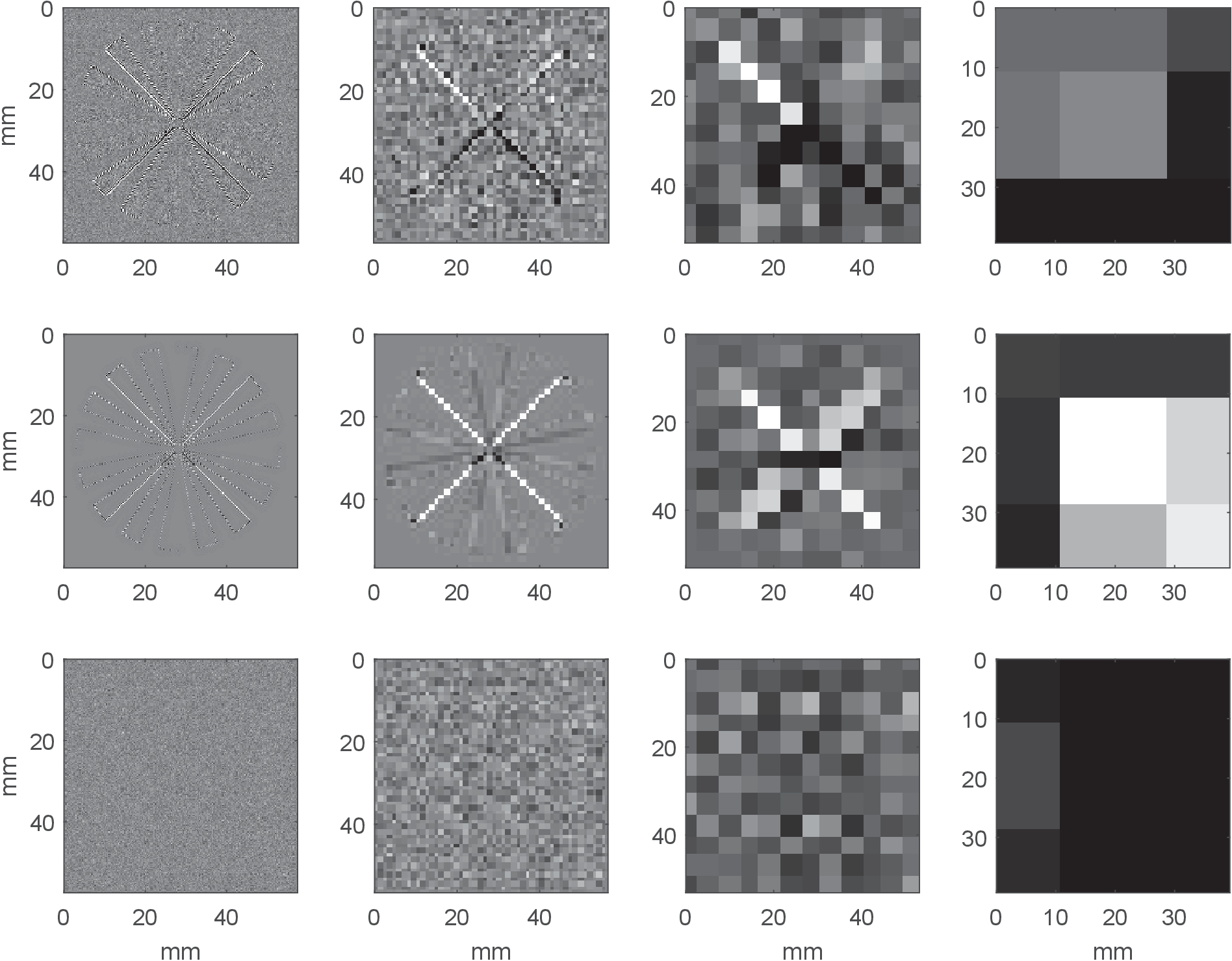}}

\caption{A sample of the nodes in the wavelet packets of images in figure \ref{fig:sig_orig} used for detectability indexes calculations. Columns 1 to 4 correspond to wavelet packets of $L=1$, $L=3$, $L=5$ and $L=7$ levels respectively. Rows 1, 2 and 3 present nodes $(L,3\times4^{L-1})$ of signal $Wg_s$, template $WHf_s$ and noise $Wg_{as}$ respectively}
\label{fig:test_signals}
\end{figure}

A sample of the test signals produced in this work is shown in figure \ref{fig:test_signals}. Columns 1, 2, 3 and 4 correspond to wavelet packets of $L=1, 3, 5, 7$ levels respectively. In can be seen how the size of the images (in pixels) decrease as the number of levels increase. The first row of this image represents nodes of the output image $Wg_s(L,k_0)$. Nodes for the template $WHf_s(L,k_0)$ and the noise $Wg_{as}(L,k_0)$ are presented in rows 2 and 3 respectively. In all cases, the node represented is $k_0=(L,3\times4^{L-1})$ that is located in the high frequency area of the image. Detectability indexes are obtained as scalar products of nodes in rows 2 and 1 (equation \ref{eq:npw1}) and scalar products of nodes in rows 2 and 3 (equation \ref{eq:npw2}).

\section{Results}\label{sec:results}
\begin{figure}
\centering
\subfigure[RQA 3, 1.79 $\mu Gy$]
{\includegraphics[width=0.48\textwidth]{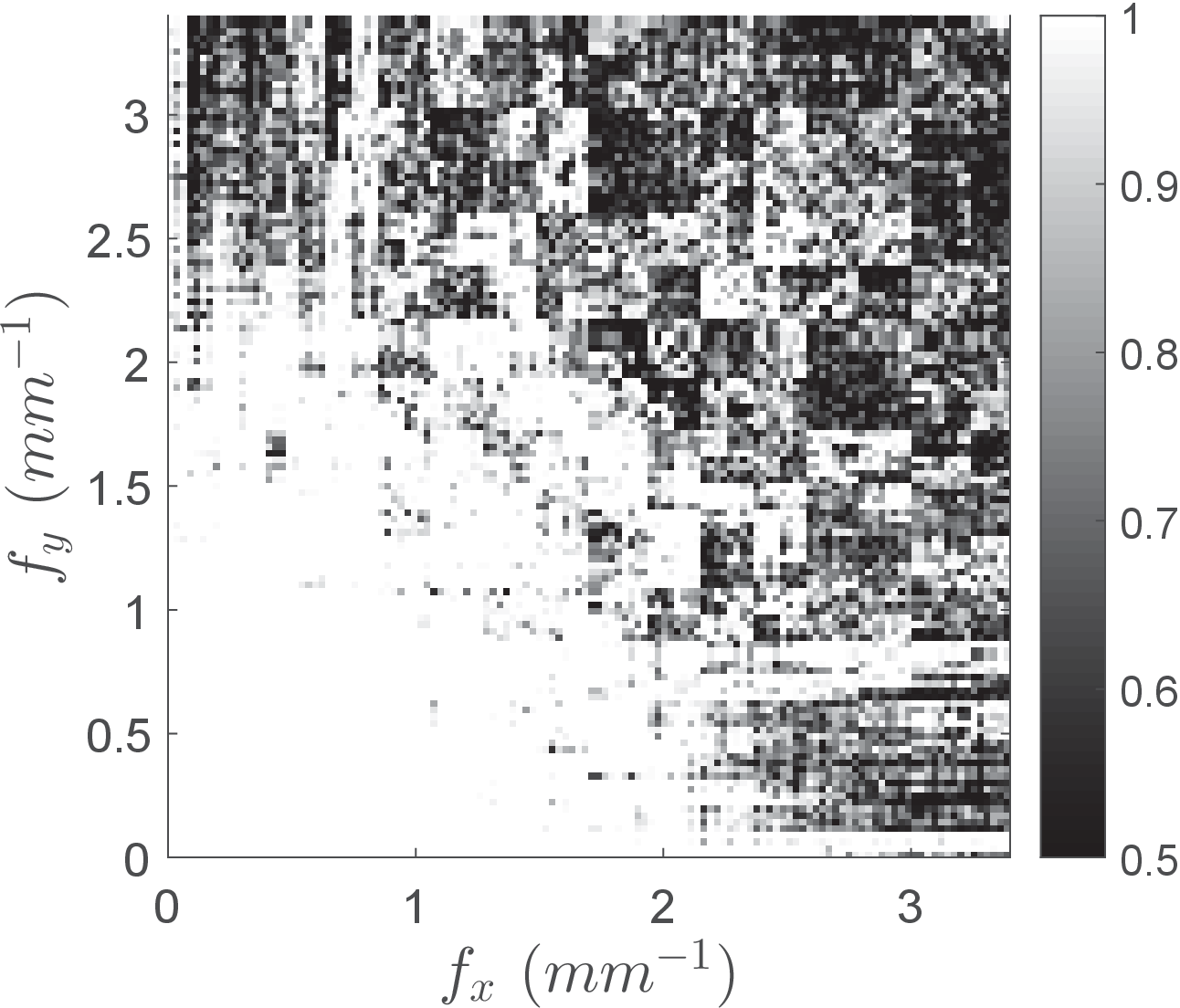}\label{fig:AUC_map_r3_02_sym16_L7}}
\subfigure[RQA 3, 17.9 $\mu Gy$]
{\includegraphics[width=0.48\textwidth]{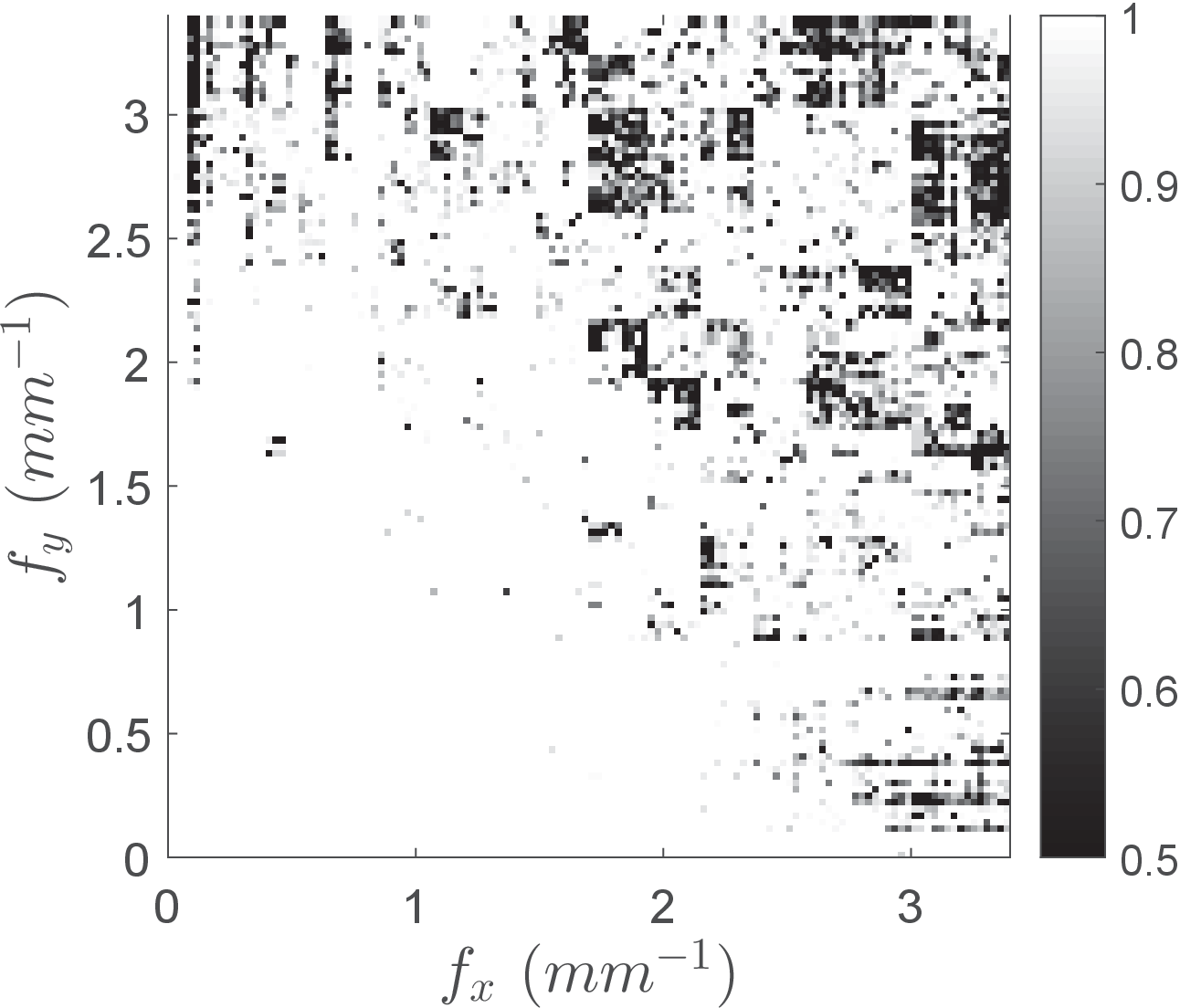}\label{fig:AUC_map_r3_20_sym16_L7}}
\subfigure[RQA 5, 1.76 $\mu Gy$]
{\includegraphics[width=0.48\textwidth]{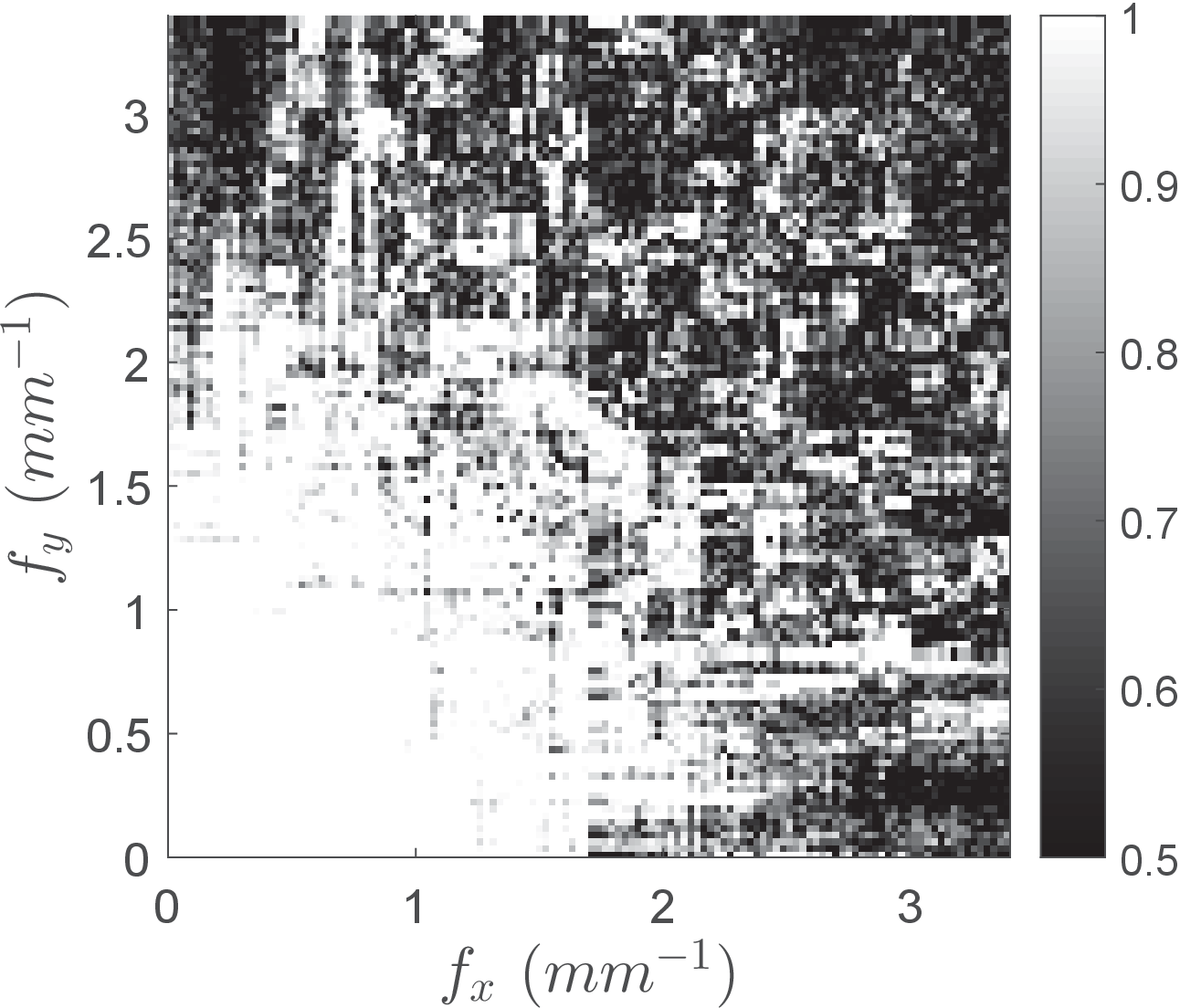}\label{fig:AUC_map_r5_1p8_sym16_L7}}
\subfigure[RQA 5, 7.99 $\mu Gy$]
{\includegraphics[width=0.48\textwidth]{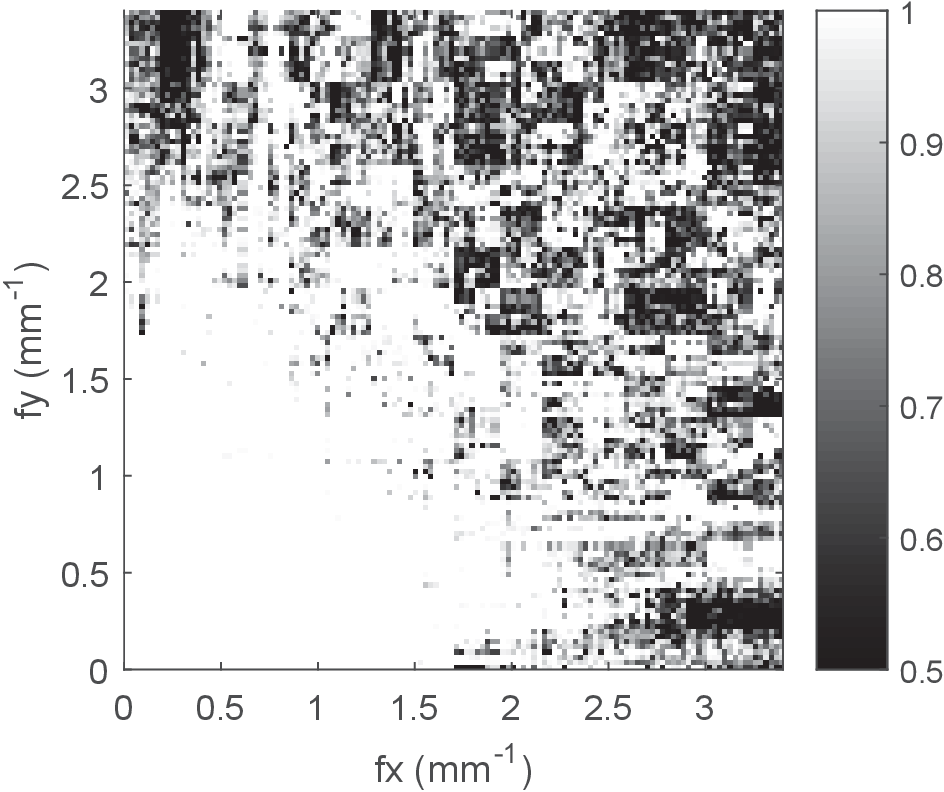}\label{fig:AUC_map_r5_8p1_sym16_L7}}
\caption{AUC maps presenting the area under the ROC curve for the different nodes of the wavelet packet of the star-bar object. AUC maps for two beam qualities and four entrance air KERMA are shown. A wavelet packet using sym16 wavelets and 7 levels decomposition has been used.}
\label{fig:AUC_maps_sym16_L7}
\end{figure}

Figure \ref{fig:AUC_maps_sym16_L7} shows AUC values for each node of the wavelet packets of the star-bar images. The wavelet packets use a sym16 wavelet and carry out 7-levels decomposition. AUC for each node is presented in the frequency coordinates assigned to the node in figure \ref{fig:freccont7lev}. This assignment gives rise to AUC maps in figures \ref{fig:AUC_map_r3_02_sym16_L7}, \ref{fig:AUC_map_r3_20_sym16_L7}, \ref{fig:AUC_map_r5_1p8_sym16_L7} and \ref{fig:AUC_map_r5_8p1_sym16_L7} that correspond to the different beam qualities and entrance air KERMA used in this work.

For low frequency nodes, figure \ref{fig:AUC_maps_sym16_L7} shows that AUC values are close to 1 in all images. However, as the frequency increases, along any directional axis the AUC values tend to decrease. Also, highest values of AUC are found in images acquired with higher doses for both beam qualities.

\begin{figure}
\centering
{\includegraphics[width=0.63\textwidth]{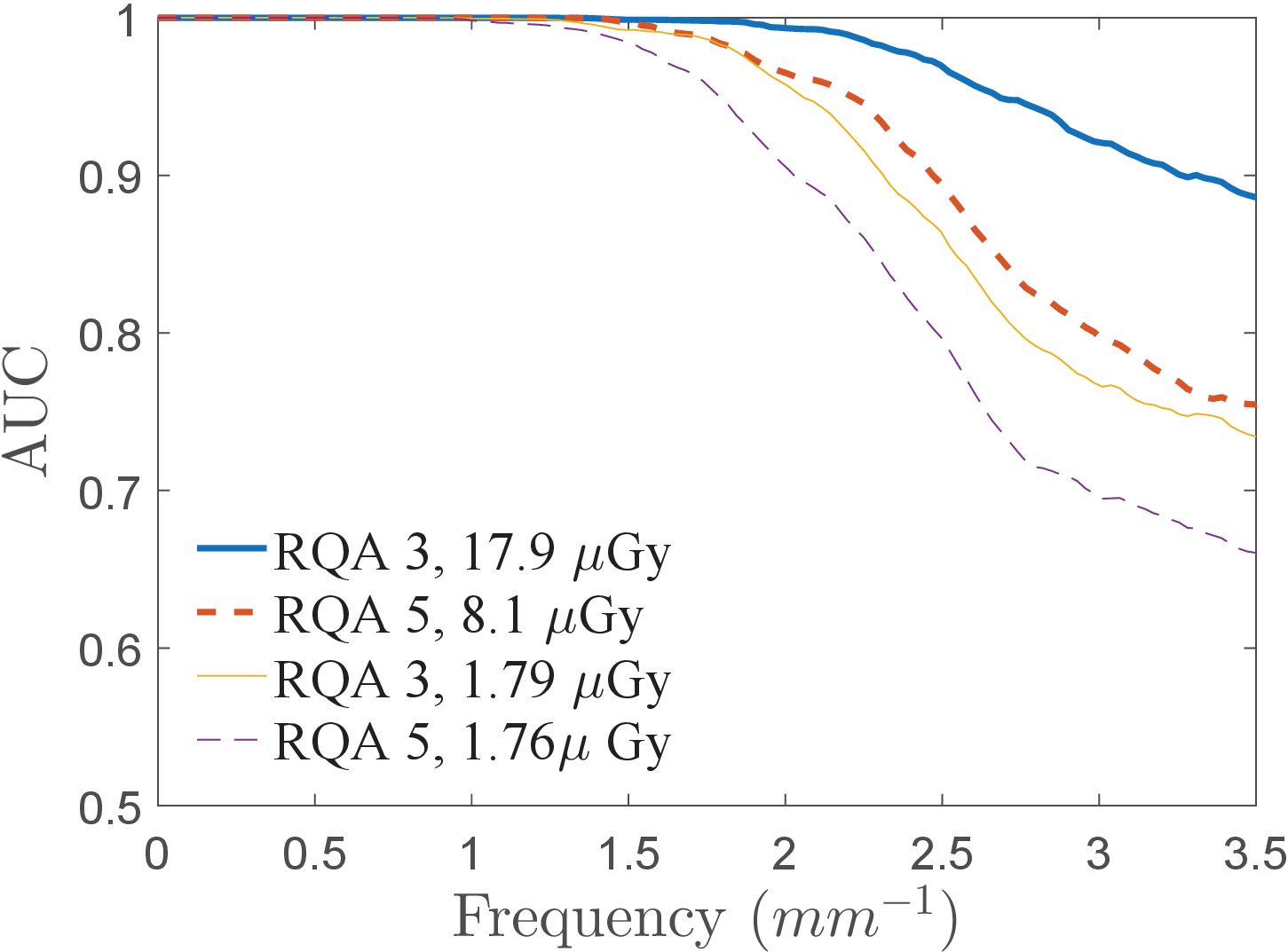}}
\caption{AUC curves obtained by radially averaging the AUC maps in figure \ref{fig:AUC_maps_sym16_L7}.}
\label{fig:AUC_curve_sym16_L7}
\end{figure}

Figure \ref{fig:AUC_curve_sym16_L7} shows AUC curves obtained by radially averaging the AUC maps presented in figure \ref{fig:AUC_maps_sym16_L7}. These curves have also been smoothed by a moving average filter with a length of 0.3 $mm^{-1}$. The advantage of this representation with respect to AUC maps is that quality of images can be more easily compared. For the test object and the assessment method used in this work, the best results on detectability are reached by the lowest beam qualities and the highest entrance air KERMAs. 

\begin{figure}
\centering
\subfigure[Haar, L=7]
{\includegraphics[width=0.49\textwidth]{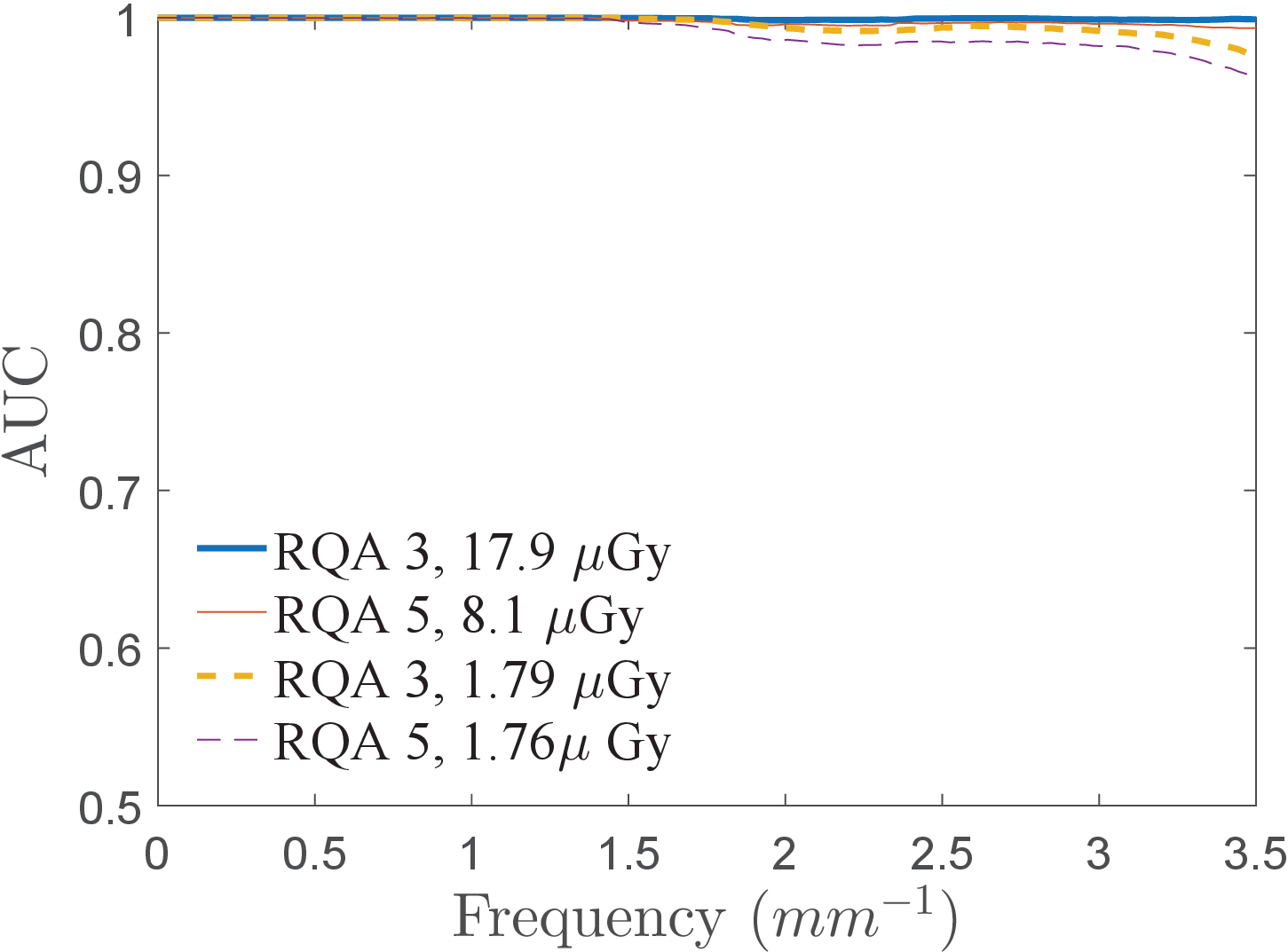}\label{fig:AUC_curve_haar_L7}}
\subfigure[Sym16, L=6]
{\includegraphics[width=0.49\textwidth]{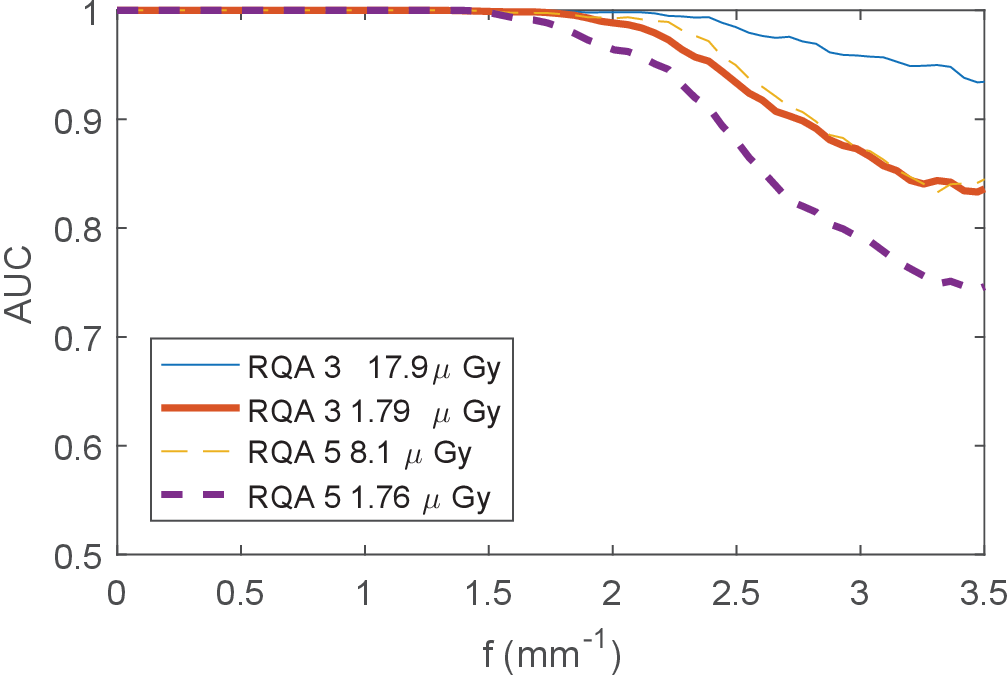}\label{fig:AUC_curve_sym16_L6}}
\caption{Effect of using a different wavelet and a different number of decomposition levels on the AUC results.}
\label{fig:AUC_curves_others}
\end{figure}

The results obtained with this method will depend on the linear transformation $W$ used. Two fundamental parameters of the wavelet packet used as $W$ in this work are the wavelet selected and the number of levels in the decomposition. How these parameters affect detectability results can be seen in figure \ref{fig:AUC_curves_others}. Figure \ref{fig:AUC_curve_haar_L7} shows the effect of changing from a sym16 wavelet to a Haar wavelet, and figure \ref{fig:AUC_curve_sym16_L6} shows how results are affected when the number of levels is changed from 7 to 6. These differences stress that the same wavelet packet, set up with the same parameters, must be used to compare different equipment or different exposure techniques.

\section{Discussion}
Quality control test based in specific tasks are expected to be more predictive of clinical performance of imaging systems that traditional specification-based tests of physical parameters like large-scale transfer function, spatial resolution and noise \cite{Samei2019,ICRU1995}.

A large number of task-based tests are being developed. Among them, tasks designed for detectability are of great interest since detectability shows a strong relationship with dose and is one of the most important characteristics defining image quality.

There are a large number of detectability tests that can be performed since there are a large number of signals that can be used for detectability. For instance, by using linear transforms, detectability can be studied in different images than those that have been acquired. In this regard equation \ref{eq:likeratio2} describes how likelihood ratios are modified for an ideal Bayesian observer when linear transforms are used. 

In this work, a particular linear transformation of the data has converted a large-size and high-contrast image in a set of small images with varying contrasts. Among the advantages of using a wavelet packet as the linear transform one can find that signals to use for detectability are two-dimensional and conform a set of signal of several levels of contrast and different frequency contents.

Using a star-bar phantom image as the original image or node $(0,0)$ of the wavelet packet has the advantage that the object has large high frequency contents \cite{Gonzalez-Lopez2015} and these high frequency contents are found in all spatial directions. This is particularly important for a transform like the wavelet packet that analyses frequency contents in all spatial directions too.

Detectability analysis has been carried out through the whole 2D frequency spectrum of the images in the AUC maps of figure \ref{fig:AUC_maps_sym16_L7}. The resolution of these maps is determined by the number of levels in the wavelet packet, since by increasing the number of levels the frequency band for the nodes is reduced (see figure \ref{fig:wltx2d}).

Increasing the number of levels in the wavelet packet makes that nodes resulting from the transformation have a worse signal to noise ratio due to each new level carries out two factor 2 downsampling that reduces the number of pixels of the node to $1/4$ of the number of pixels in the preceding level node. In this way, by increasing the number of levels, signals with a worse detectability are produced. This explains the different AUC values shown in figures \ref{fig:AUC_curve_sym16_L7} and \ref{fig:AUC_curve_sym16_L6}.

Another important parameter of the wavelet packet is the wavelet used in the transformation. Figure \ref{fig:fils} shows the decomposition filters for two different wavelets. It has been mentioned that the sharper shape of sym16 filters transfer the frequency contents more efficiently between nodes. In the case of the Haar filters, a large fraction of low frequency components is transferred to the high frequency part of the spectrum and a large fraction of high frequency components is transferred to the low frequency components of the output nodes. Because of this, detectability remains high at high frequencies when the Haar wavelet is used (figure \ref{fig:AUC_curve_haar_L7}).
\section{Conclusion}
A task-based quality-control test for detectability assessment of an x-ray imaging system has been presented. From a star-bar pattern object a set of test images is generated by applying a wavelet packet $W$ to the different terms in equation \ref{eq:transfer}. Then matching between the template $WHf_s$ and the transformed output image $Wg$ (equations \ref{eq:npw1} and \ref{eq:npw2}) are calculated to produce detectability indexes for a NPW observer. Finally, ROC analysis is applied to evaluate detectability performance of the system using images acquired with different doses and different beam qualities.

Using a wavelet packet on a star-bar object allows studying detectability of the imaging system in the whole frequency space by means of the AUC maps. The results can be used to compare different equipment and different acquisition techniques. In particular, the ability of the method for distinguishing different dose levels and beam qualities can be used to study the image quality-dose relationship.

Linear transforms allows exploring detectability of an imaging system on a wide number of scenarios by simply adapting the transform characteristics to the target of study. In this work, the target consisted in using 2D images as signals to detect and producing a set of signals with largely varying contrast and frequency contents.

\section*{Acknowledgement}
This work has been supported by the Comunidad Aut\'onoma de la Regi\'on de Murcia, Spain (Ref. 20861/PI/18) through the call for grants to projects for the development of scientific and technical research by competitive groups, included in the Regional Program for the Promotion of Scientific and Technical Research (Action Plan 2018) of the Fundaci\'on S\'eneca-Agencia de Ciencia y Tecnolog\'ia de la Regi\'on de Murcia.

\section*{References}
\addcontentsline{toc}{section}{\numberline{}References}
\vspace*{-20mm}

\bibliography{./ctt3}      
\bibliographystyle{./medphy.bst}    

\end{document}